\documentclass[sigconf]{acmart}
\AtBeginDocument{%
  }

\setcopyright{none}  
\acmConference[Preprint]{}  
\acmYear{}  
\copyrightyear{}  
\acmDOI{}  
\acmISBN{}  

\usepackage{threeparttable}

\begin{document}

\title{Enhancing reliability in AI inference services: An empirical study on real production incidents}

\author{Bhala Ranganathan}
\affiliation{%
  \institution{Microsoft}  
  \country{Redmond, WA, USA}
  }
\orcid{0009-0001-2744-9655}
\email{bheroder@microsoft.com}

\author{Mickey Zhang}
\affiliation{%
  \institution{Microsoft}  
  \country{Redmond, WA, USA}
}
\email{ruizh@microsoft.com}

\author{Kai Wu}
\affiliation{%
  \institution{Microsoft}
  \country{Redmond, WA, USA}
}
\email{kai.wu@microsoft.com}
\authornote{Contributed to incident dataset collection and initial discussions}

\renewcommand{\shortauthors}{Ranganathan et al.}

\begin{abstract}
Hyperscale large language model (LLM) inference places extraordinary demands on cloud systems, where even brief failures can translate into significant user and business impact. To better understand and mitigate these risks, we present one of the first provider-internal, practice-based analysis of LLM inference incidents. We developed a taxonomy and methodology grounded in a year of operational experience, validating it on 156 high-severity incidents, and conducted a focused quantitative study of Apr–Jun 2025 to ensure recency and relevance. Our approach achieves high labeling consistency (Cohen’s K $\approx$ 0.89), identifies dominant failure modes (in our dataset $\approx$60\% inference engine failures, within that category $\approx$40\% timeouts), and surfaces mitigation levers ($\approx$74\% auto‑detected; $\approx$28\% required hotfix). Beyond hotfixes, many incidents were mitigated via traffic routing, node rebalancing, or capacity increase policies, indicating further automation opportunities. We also show how the taxonomy guided targeted strategies such as connection liveness, GPU capacity-aware routing, and per-endpoint isolation and reduced incident impact and accelerated recovery. Finally, we contribute a practitioner-oriented adoption checklist that enables others to replicate our taxonomy, analysis, and automation opportunities in their own systems. This study demonstrates how systematic, empirically grounded analysis of inference operations can drive more reliable and cost-efficient LLM serving at scale.
\end{abstract}



\maketitle

\section{INTRODUCTION}
\subsection{Motivation}
The rapid adoption of large language models (LLMs)  \cite{touvron2023llama2openfoundation} \cite{grattafiori2024llama3herdmodels} \cite{llama} in production introduces new reliability challenges at hyperscale. \textbf{LLM serving must deliver low latency, highly available responses across diverse workloads, and as usage and model complexity grow, incidents carry significant customer impact and engineering cost}. Improving resilience starts with understanding the production incident lifecycle: identifying recurring root causes enables proactive detection and remediation, while quantifying effort across detection, diagnosis, and mitigation exposes process bottlenecks and guides automation. \textbf{We present an empirical study of 156 high-impact production incidents, leveraging a year’s worth of operational learnings and focusing our quantitative analysis on the most recent three months to ensure data relevance and minimize confounding factors}. We characterize root causes, assess mitigation efficacy, and distill operational lessons. Our findings complement prior cloud reliability work by centering on customer facing LLM serving, calibrate claims to this context, and surface insights likely to generalize to AIOps, Site Reliability Engineering (SRE), and AI platform governance.

\subsection{Research/practitioner questions}
Reliability incidents in cloud-scale \textbf{LLM inference systems can disrupt thousands of customer requests, resulting in lost revenue, increased support costs, and reputational risk.} By identifying dominant failure modes and evaluating effective mitigations, we enable targeted investments that reduce incident frequency and duration translating into cost savings and improved customer satisfaction. Quantifying these impacts helps justify engineering priorities and demonstrates the tangible business value of operational resilience.
\begin{itemize}
    \item \textbf{Q1}: What failure modes dominate high impact incidents in LLM serving, and what are their relative shares?
    \item \textbf{Q2}: Which mitigations are most frequently effective and where do automation gaps remain?
    \item \textbf{Q3}: How do incident and inference request lifecycle metrics evolve over the study window?
\end{itemize}

\subsection{Key contributions}
\begin{itemize}
\item \textbf{A validated taxonomy and labeling rulebook for LLM inference incidents}: We introduce a four-way taxonomy (infrastructure failures, model configuration failures, inference engine failures, operational failures) and apply it to 156 high-severity incidents, achieving strong inter-rater reliability (Cohen’s kappa \cite{cohen1960} $\approx$ 0.89). See \autoref{tab:table1}, Section 3.2.
\item \textbf{A quantified landscape of failure modes, error codes, and lifecycle metrics applying the taxonomy}: We provide a detailed breakdown of dominant incident types, error code distributions, and both incident and request lifecycle metrics in our dataset as summarized in \autoref{tab:headline-stats}, \autoref{tab:analysis-summary}, \autoref{tab:mitigation-summary}, \autoref{tab:Inference request lifecycle metrics}
\item \textbf{Taxonomy guided cause→mitigation→outcome mapping with directional evidence for key interventions}: We map root causes to mitigation actions and operational outcomes. See \autoref{tab:cause-mitigation-outcome} and Section 5.4.
\item \textbf{A practitioner focused adoption plan and generalization checklist}: We distill actionable lessons such as the need for more AIOps automation and intelligent failover as rapid recovery actions for auto healing and operational tooling. See Section 7.1 and Section 5.5.
\end{itemize}

\subsection{Practitioner quick takeaways}
These recommendations are based on a year of operational experience, with empirical analysis focused on the most recent, representative incidents. Also see Section 7 for adoption checklist and future research directions.
\begin{itemize}
  \item \textbf{Adopt and share incident taxonomy/rulebook:} Accelerates incident analysis and cross-team learning.
  \item \textbf{GPU efficiency + capacity planning:} Tune request batching, cap pending tokens \cite{10.5555/3295222.3295349}; plan proactive capacity.
  \item \textbf{Intelligent failover + GPU node rebalancing \cite{10.1145/2890784}:} Automate health probes and traffic routing to speed up recovery.
  \item \textbf{Automate mitigation playbooks:} Self-healing to cut human intervention.
  \item \textbf{Adaptive AIOps monitoring (incl.\ low-traffic endpoints):} Use dynamic thresholds/anomaly detection to catch silent degradations.
\end{itemize}

\subsection{System architecture overview}
\begin{figure}[ht]
\centering
\includegraphics[width=\columnwidth]{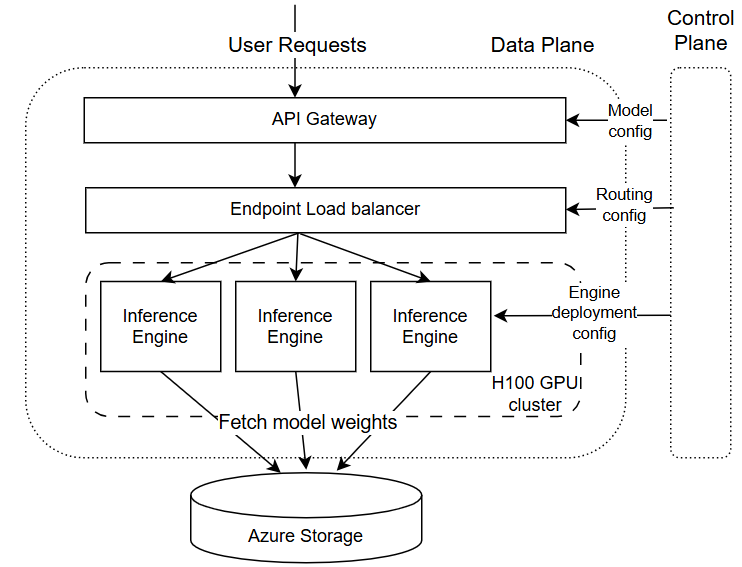}
\caption{An illustration of the LLM serving architecture}
\Description{}
\label{fig_arch}
\end{figure}

At a high level, our LLM serving system (\autoref{fig_arch}) is composed of control plane and data plane components. The control plane manages configurations across different layers, including model-specific settings, endpoint traffic routing, load balancing policies, and engine deployment parameters. The data plane provides components for scheduling and serving inference requests, enabling flexible and high performance model execution. AI models are deployed as containerized services on GPU clusters managed by Kubernetes, with both off-the-shelf and custom variants in production. The data plane components include:
\begin{itemize}
\item \textbf{API gateway}: Receives inference requests via REST or gRPC \cite{grpc-github}.
\item \textbf{Endpoint load balancer}: Applies health probes and GPU capacity-aware policies to direct traffic. Also aggregates requests, manages request batching, and applies token \cite{10.5555/3295222.3295349} caps.
\item \textbf{Inference engine}: Loads model weights, executes requests, manages KV cache \cite{305212} and sampling parameters. These inference engines run on GPU clusters managed by Kubernetes.
\end{itemize}

\noindent This cloud-scale inference platform analyzed in this study serves a diverse set of production AI models, primarily LLMs for text generation and classification (e.g., Llama-2 \cite{touvron2023llama2openfoundation}), as well as vision and speech models (e.g., Llama-3 \cite{grattafiori2024llama3herdmodels}) for multi-modal tasks. For this study, we consider reasoning models as those designed for multi‑step logical or mathematical reasoning (e.g., models targeting multi‑step reasoning benchmarks such as BIG‑bench \cite{srivastava2023beyond}). The majority of incidents in our dataset ($\approx$75\%) involved non-reasoning LLMs, while the remainder included reasoning models and select embedding models used for retrieval and ranking. \textbf{Key system characteristics include high concurrency, variable prompt lengths, and support for streaming inference, which collectively drive unique reliability challenges compared to traditional cloud workloads}. Such architecture enables rapid failover, supports connection liveness strategies for long-running requests, and robust error handling, as evidenced by our mitigation analysis (Section 5).\\

\noindent The remainder of this paper is structured as follows: We discuss related work (Section 2), define a four-way incident taxonomy and labeling rulebook (Section 3), apply it to a real production corpus to quantify failure modes and error codes (Section 4), evaluate mitigation actions and automation opportunities (Section 5), discusses a few case studies (Section 6) and translate findings into a practitioner checklist (Section 7). Finally we discuss the limitations of our analysis (Section 8), followed by conclusion (Section 9).

\section{RELATED WORK}
\label{sec:RELATED WORK}
The operational reliability of large scale AI inference systems has become a critical area of study as demand for generative AI services continues to grow. We position our results alongside cloud reliability (e.g., deployment monitoring and automated failover), inference serving optimization (e.g., request batching, KV cache efficiency), and intelligent incident management (AIOps). \textbf{Our focus is on customer facing LLM serving at scale, complementing broader cloud reliability studies by detailing model/engine specific failure patterns}.

\subsection{Cloud services reliability}

Reliability in cloud computing has been a central concern as services scale to meet global demand. Ensuring high availability and resilience in cloud services has long been a foundational goal for hyperscale platforms. Traditional approaches have emphasized redundancy, failover mechanisms, and reactive monitoring. Ghosh et al. (2022) \cite{10.1145/3542929.3563482} uncover interesting insights by a novel multi dimensional analysis that correlates different troubleshooting stages, and provide guidance on how to tackle complex incidents through automation or testing at different granularity.  However, recent research has shifted toward predictive and AI driven strategies. Microsoft’s Azure platform, for instance, has adopted a multi pronged approach to reliability that includes proactive monitoring, automated failover, and service level objective (SLO) enforcement. Further, Srinivas et al. (2024) \cite{srinivas2024intelligent} propose a deep learning based framework that recommends monitors for cloud services based on their service properties to ensure high availability and reliability and derive key insights on the major classes of monitors employed by cloud services at Microsoft.  Kokolis et al. (2025) \cite{kokolis2025revisitingreliabilitylargescalemachine} identify key machine learning workload properties introduce a taxonomy of failures and key reliability metrics analyzing 11 months of data emphasizing the need for flexible, workload agnostic, and reliability aware infrastructure. 

\subsection{AI inference engine reliability}

AI inference systems such as vLLM \cite{vllm} introduce unique reliability challenges due to their sensitivity to model behavior, data distribution, and infrastructure variability.  In operational environments, inference engines are monitored using fine grained metrics. These metrics are critical for diagnosing performance bottlenecks and ensuring Service Level Agreement (SLA) compliance.  The use of synthetic inference tests and container validation pipelines further supports reliability practices, enabling early detection of regressions before production deployment. Several papers have studied and improved the performance of LLM serving systems. Yu et al. (2022) \cite{280922} introduces ORCA system which schedules execution at the granularity of iteration and can significantly outperform NVIDIA FasterTransformer in terms of both latency and throughput. Agrawal et al. (2024) \cite{agrawal2024tamingthroughputlatencytradeoffllm} propose techniques like chunked-prefills that yield significant improvements in inference performance across models and hardware under tail latency constraints. Industry articles by Nvidia \cite{Nvidia}, Databricks \cite{Databricks} present comprehensive benchmarks evaluating the performance of LLM serving systems. A few other studies which try to improve the efficiency of LLM serving by targeting varied set of problems are \cite{kwon2023efficientmemorymanagementlarge}, \cite{ miao2023spotserveservinggenerativelarge}, \cite{ wu2024fastdistributedinferenceserving}, \cite{hu2025deepserveserverlesslargelanguage}, \cite{298681}, \cite{305212}. Also, unlike public outage studies \cite{Chu_2025} that analyze cross provider incident feeds and open incident reports, our study examines an internal corpus of high severity inference related incidents.

\subsection{Intelligent incident management systems}
Incident management in cloud-scale systems has evolved from manual triage to increasingly automated and intelligent workflows. Traditional systems rely on static troubleshooting guides and reactive escalation paths. However, recent advancements incorporate AI agents for root cause analysis, incident enrichment, and automated remediation.  Bansal et al. (2020) \cite{10.1145/3377813.3381353} introduce DeCaf, a system for automated diagnosis and triaging of KPI issues using service logs. It uses machine learning along with pattern mining to help service owners automatically root cause and triage performance issues.  Yu et al. (2025) \cite{yu2025triangle} proposes Triangle, an end-to-end incident triage system based on a Multi-LLM-Agent framework.  Chen et al. (2020) \cite{10.1145/3368089.3417055} conduct a comprehensive empirical study spanning over two years of incident management practices at Microsoft and investigate the underlying reasons from the perspective of cloud system design and operations. They present IcM BRAIN, an AIOps framework towards intelligent incident management, and show its practical benefits conveyed to the cloud services of Microsoft. Saha et al. (2022) \cite{10.1145/3510457.3513030} present Incident Causation Analysis (ICA) and the downstream Incident Search and Retrieval based Root Cause Analysis (RCA) pipeline, built at Salesforce, over 2000 documented cloud service incident investigations collected over a few years.\\

\noindent While the above related work provides essential context for maintaining hyperscale platforms through cloud resiliency and LLM optimization techniques, \textbf{to our knowledge, this is among the first practice‑based, end‑to‑end analysis focused specifically on reliability issues in customer facing LLM serving at hyperscale, grounded in production operations.} This specific focus enables the detailed exploration of incident root causes and mitigation strategies, such as those related to inference engine failures, model configuration issues, and GPU utilization, which are distinct from general cloud challenges and inform the actionable insights and future work presented throughout the study.

\section{METHODOLOGY}
\label{sec:METHODOLOGY}
To understand operational challenges and resilience patterns in large-scale AI inference systems, we analyzed high severity incidents over a focused three month period (April–June 2025). \textbf{Our taxonomy and key lessons are distilled from a full year of running the service, with the three-month dataset chosen for its high signal and relevance.} This window avoids seasonal and architectural noise, aligns with major deployments, and ensures that our findings reflect the current state of the system. Our methodology combines quantitative and qualitative approaches to capture the full incident lifecycle.

\subsection{Production incident data collection}

While automation can streamline data processing, we manually curated data in Excel from multiple sources to ensure rigor and accuracy in all correlation and association analysis. For each day in the study period, we recorded incident details and other relevant variables based on operational logs and deployment schedule. Data entry was cross checked with engineers to ensure accuracy.

\begin{itemize}
    \item \textbf{Incident Management System (ICM)}: Used to extract metadata for all severity 2 and 2.5 incidents, including timestamps, severity, impacted endpoints, and resolution notes.
    \item \textbf{Postmortem reports}: Provided detailed narratives of root causes, mitigation steps, and follow up actions.
    \item \textbf{Runtime logs and dashboards}: Offered telemetry on engine health, latency, error rates, and capacity utilization.
    \item \textbf{Designated Responsible Individual (DRI) playbooks and tools}: Helped contextualize the operational workflows and automation used during incident response.
\end{itemize}

Our incident management system rates severity from 0 (most critical) to 4 (least). Severity 2.5 is an internal label for incidents as impactful as severity 2 but resolved during business hours. \textbf{We focus on severity 2 \& 2.5 to target customer impacting incidents; lower severities are excluded to avoid diluting analysis with low signal events}. We report how this scope may bias trends in Section 8. We analyze all 156 high severity incidents during the study period i.e. 129 severity 2 and 27 severity 2.5 incidents.

\subsection{Root cause categorization}
\label{subsec:Root cause categorization}
Each incident was manually reviewed and assigned to one of the four root cause categories (\autoref{tab:table1}). In cases where an incident involves multiple contributing factors, \textbf{we resolve classification ambiguity by selecting the one that appears earliest in the incident summary}. This taxonomy was iteratively refined through cross validation with incident reviewers and inference serving team engineers.

\begin{table}[ht]
\caption{Taxonomy of incidents \label{tab:table1}}
\centering
\small
\renewcommand{\arraystretch}{1.2}
\setlength{\tabcolsep}{6pt}      
\begin{tabular}{p{0.2\columnwidth} p{0.7\columnwidth}}
\toprule
 \textbf{Category}& \textbf{Rulebook scenarios}\\
 \midrule
 Infrastructure failures& Compute capacity constraints, delayed compute allocation, misconfigured deployment templates, bad nodes, redis \cite{redis-github}, MPI ring \cite{mpi41}, sidecar, istio \cite{istio}, node scheduling and storage. \\
 \midrule
 Model configuration failures&Invalid weights, missing or misconfigured headers and invalid output.\\
 \midrule
 Inference engine failures&Resource exhaustion, inference timeouts, latency spikes, KV cache errors, crashes, input fetch, constraint sampling, memory leaks.\\
 \midrule
 Operational failures&Delayed detection due to low-traffic, misconfigured alerts, or missing telemetry, error code translation, imbalanced traffic routing and rollout issues.\\
 \bottomrule
\end{tabular}
\end{table}

\subsection{Labeling procedure and inter-rater reliability} 
Incidents were the unit of analysis. The dataset comprises 156 incidents. Two raters independently labeled a pilot subset ($\approx$20\%) using a draft rulebook covering four root cause categories (Infrastructure failures, Model configuration failures, Inference engine failures, Operational failures). Disagreements were reconciled, decision rules were refined, and the rulebook was updated. See \autoref{tab:table1}. Then, five engineers independently labeled $\approx$31 incidents each. Subsequently, to assess labeling reliability, we selected 31 incidents (random $\approx$6 per rater) and independently assigned labels using our rulebook. Agreement between the initial and reassigned labels was $\approx$90\%. \textbf{Cohen’s kappa \cite{cohen1960}, which adjusts for chance agreement, was 0.89 (95\% CI [0.75, 1.00])}, indicating almost perfect consistency in the application of labeling rules. Remaining disagreements were resolved by consensus, and the adjudicated labels were used for all analysis.

\subsection{Empirical study approach}
To systematically understand the dynamics of high severity incidents, we examined not only their root causes but also the operational pathways through which they were detected, diagnosed, and resolved. Our objective was to uncover patterns that inform more effective incident response and prevention strategies. In practice, incident handling follows a four phase progression:
\begin{itemize}
    \item \textbf{Detection:} This is where the issue is surfaced either through automated monitoring systems or customer reports
    \item \textbf{Diagnosis:} This phase involves root cause analysis by the team
    \item \textbf{Mitigation:} The phase where corrective actions are implemented to restore service health and prevent recurrence.
    \item \textbf{Postmortem:} A reflection of learning from handling this incident.
\end{itemize}

To capture the nuances of this lifecycle, we followed the six operational dimensions that influence the effectiveness of incident management \cite{10.1145/3542929.3563482}. These dimensions detailed in \autoref{tab:table4} span detection latency, escalation clarity, tooling support, mitigation complexity, documentation quality, and cross-team coordination. Together, they provide a structured lens through which to evaluate incident response workflows and identify opportunities for automation, process refinement, and knowledge sharing.

\begin{table}[ht]
\caption{Factors considered \label{tab:table4}}
\centering
\small
\renewcommand{\arraystretch}{1.2}
\setlength{\tabcolsep}{6pt}      
\begin{tabular}{p{0.3\columnwidth} p{0.6\columnwidth}}
\toprule
\textbf{Study factor}& \textbf{Description}\\
\midrule
Root cause& What issue caused the incident?\\
\midrule
 Mitigation steps&What steps were performed to restore service health?\\
 \midrule
 Detection failure&Why did monitoring not detect the incident?\\
 \midrule
 Mitigation failure&What challenges delayed incident mitigation?\\
 \midrule
 Automation opportunities& What automation can help improve service resilience?\\
 \midrule
 Lessons learned&What lessons were learned about the service’s behavior?\\
 \bottomrule
\end{tabular}
\end{table}

\section{INCIDENT analysis}
\label{sec:INCIDENT ANALYSIS}
We now validate the proposed taxonomy (\autoref{tab:table1}) by applying it to 156 high‑severity incidents and quantifying outcomes, thereby demonstrating the taxonomy’s operational efficacy. We address \textbf{Q1} by quantifying the incident categories A1-A4 in \autoref{tab:analysis-summary}, \textbf{Q2} by mapping mitigation actions M1-M6 in \autoref{tab:mitigation-summary} and \textbf{Q3} by analyzing lifecycle metrics \autoref{tab:Inference request lifecycle metrics} and \autoref{ttm}. 

\subsection{Incident trend}
\begin{figure}[ht]
\centering
\includegraphics[width=\columnwidth]{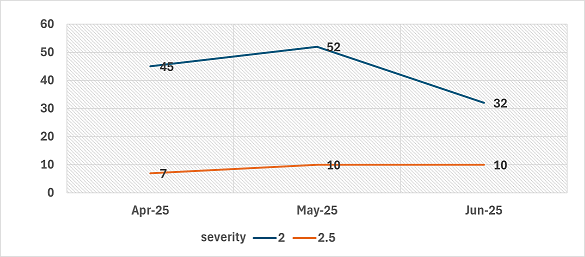}
\caption{Incident frequency by severity}
\Description{}
\label{freq}
\end{figure}

\begin{figure}[ht]
\centering
\includegraphics[width=\columnwidth]{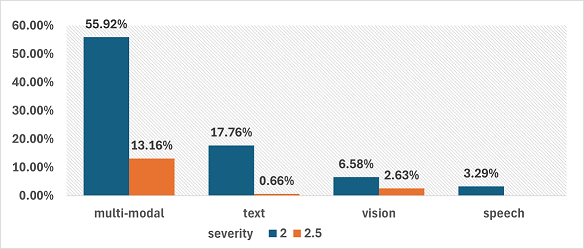}
\caption{Incidents by model modality and severity}
\Description{}
\label{freq_modality}
\end{figure}

\begin{figure}[ht]
\centering
\includegraphics[width=\columnwidth]{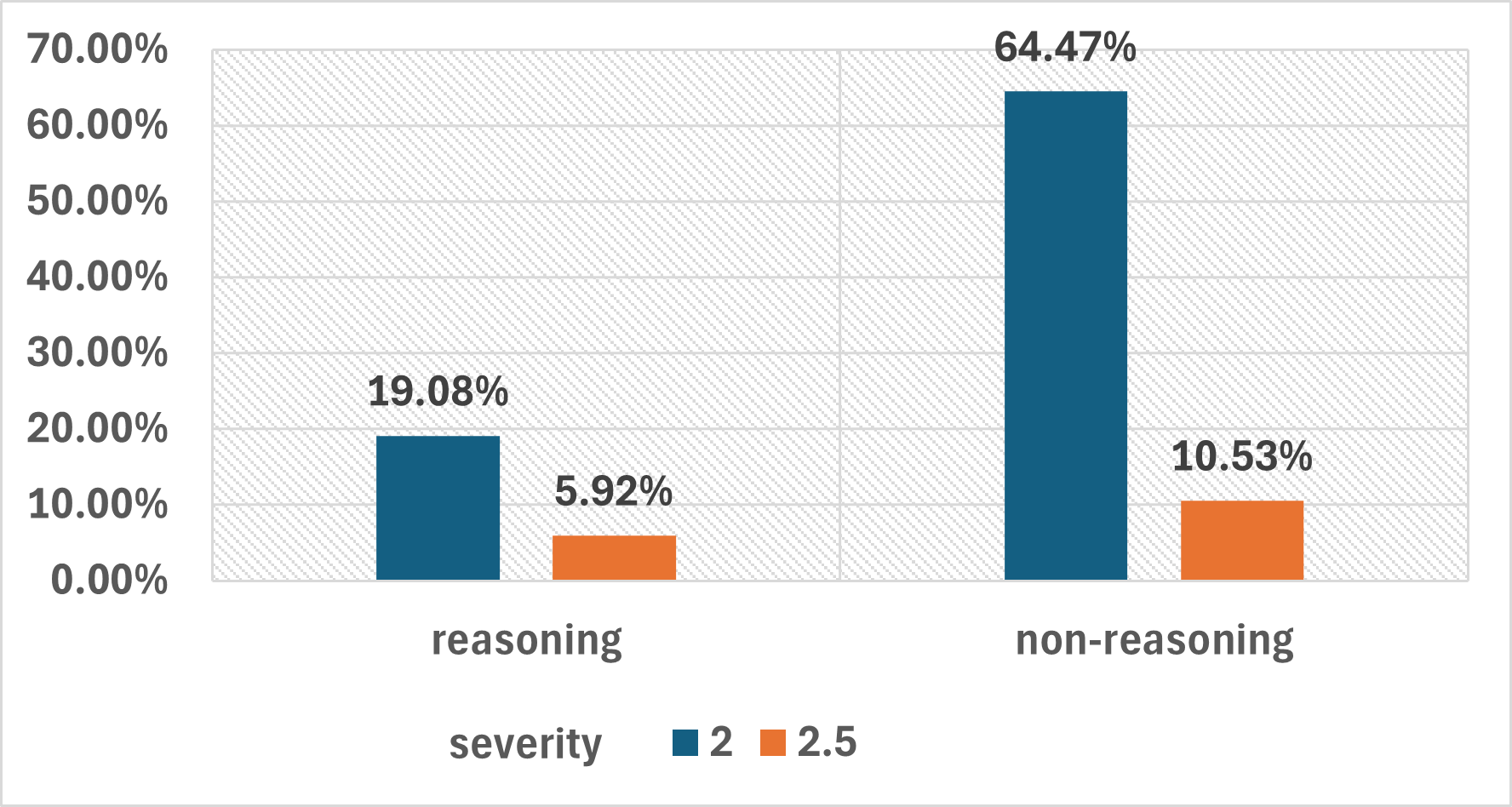}
\caption{Incidents by model family and severity}
\Description{}
\label{freq_reasoning}
\end{figure}

We begin with incident trends observed during Apr-Jun 2025 by analyzing all 156 high severity incidents. \autoref{tab:headline-stats} summarizes the main characteristics, error patterns, and operational context. Unless otherwise noted, percentages are with respect to all incidents (N=156). Key observations are:
\begin{itemize}
    \item \textbf{Incident frequency} (\autoref{freq}): May 2025 saw the highest number of severity 2 incidents
    \item \textbf{Model modality and family} (\autoref{freq_modality}, \autoref{freq_reasoning}): Non-reasoning models accounted for $\approx$75\% of incidents; most incidents were model-wide rather than modality-specific. Model diversity is reported; findings are generalized unless noted.
    \item \textbf{Error code breakdown}: Server-side errors (HTTP 5xx) dominated, with 500 errors comprising $\approx$74\% of all incidents. This indicates monitoring gaps that allowed requests to be routed even when the engine could not serve.
    \item \textbf{Scope of impact}: $\approx$96\% of incidents impacted a single endpoint, indicating effective containment.
\end{itemize}

\begin{table}[ht]
\centering
\caption{Consolidated headline statistics for high severity incidents (Apr-Jun 2025)}
\label{tab:headline-stats}
\small
\begin{threeparttable}
\begin{tabular}{p{0.7\columnwidth} p{0.1\columnwidth} p{0.1\columnwidth}}
\toprule
\textbf{What} & \textbf{Count (n/N)} & \textbf{Share}\tnote{1}\\
\midrule
Severity 2 incidents & $129/156$ & $\approx$$82.7\%$ \\
Severity 2.5 incidents & $27/156$ & $\approx$$17.3\%$ \\
\midrule
Non-reasoning models\tnote{2} & $117/156$ & $\approx$$75.0\%$ \\
Reasoning models\tnote{2} & $39/156$ & $\approx$$25.0\%$ \\
\midrule
Text modality\tnote{2} & $29/156$ & $\approx$$18.42\%$ \\
Vision modality\tnote{2} & $14/156$ & $\approx$$9.21\%$ \\
Speech modality\tnote{2} & $5/156$ & $\approx$$3.29\%$ \\
model-wide (not single modality)\tnote{2} & $108/156$ & $\approx$$69.08\%$ \\
\addlinespace[0.4em]
\multicolumn{3}{l}{\textbf{HTTP error codes}} \\
\midrule
HTTP 500 & $115/156$ & $\approx$$74\%$ \\
HTTP 408 & $16/156$ & $\approx$$10\%$ \\
HTTP 503 & $14/156$ & $\approx$$9\%$ \\
HTTP 429 & $5/156$ & $\approx$$3\%$ \\
HTTP 424 & $5/156$ & $\approx$$3\%$ \\
HTTP 504 & $1/156$ & $\approx$$1\%$ \\
\addlinespace[0.4em]
\multicolumn{3}{l}{\textbf{Scope of impact}} \\
\midrule
Incidents impacting a single endpoint & $150/156$ & $\approx$$96.2\%$ \\
Incidents impacting multiple endpoints & $6/156$ & $\approx$$3.8\%$ \\
\bottomrule
\end{tabular}
\begin{tablenotes}
\item[1] All percentages sum to 100\% per statistics grouping where N=156.
\item[2] Model diversity is reported to enhance understanding of system dynamics; findings are generalized unless noted.
\end{tablenotes}
\end{threeparttable}
\end{table}

\subsection{Infrastructure failures}
Applying our taxonomy isolated infrastructure failures as the root cause in $\approx$20\% of high severity incidents. Most infrastructure incidents (\autoref{tab:analysis-summary} A1) reflect node level failures (e.g., bad nodes, MPI ring \cite{mpi41} failures, delayed compute allocation) that surface as deployment failures or runtime instability. Node rebalancing \cite{10.1145/2890784} refers to the process of redistributing workloads from degraded or overloaded nodes to healthy ones to maintain service availability and performance. \textbf{The structured approach enabled rapid mapping from symptoms to targeted mitigations} such as node restart \cite{kubernetes_cmd} enabling rapid node rebalancing, which demonstrably shortened recovery times; buffer GPU capacity remains the limiting factor. \textbf{These findings validate the taxonomy’s utility in surfacing actionable levers for resilience} at the infrastructure layer and highlight the value of intelligent failover and capacity planning for future deployments.

\subsection{Model configuration failures }
Due to inherent limitations in HTTP protocol particularly around header size constraints and inefficient metadata handling many of our systems have transitioned to gRPC \cite{grpc-github} to enable more scalable and performant communication. \textbf{The taxonomy’s explicit separation of model configuration failures surfaced header misconfigurations and invalid outputs} (e.g., encoding/decoding errors) as the primary failure modes ($\approx$16\% of incidents; \autoref{tab:analysis-summary} A2).This clarity guided the adoption of standardized header schemas and pipeline validation, which reduced deployment related HTTP 5xx errors. By making configuration faults a first class analytic category, the method enabled precise attribution and informed the rollout of preventative controls demonstrating the practical impact of taxonomy driven analysis. See \autoref{tab:analysis-summary} (A2) and \autoref{tab:cause-mitigation-outcome} for the cause$\rightarrow$mitigation mapping. 

\subsection{Inference engine failures}
 Categorizing incidents via the taxonomy revealed inference engine failures as the most prevalent root cause ($\approx$60\% of incidents; \autoref{tab:analysis-summary} A3), with timeouts and resource exhaustion accounting for the majority. Engine incidents were largely latency and GPU capacity driven: timeouts on long-running requests and resource exhaustion under load. \textbf{This focus enabled systematic evaluation of mitigations} such as connection liveness strategies (including streaming, request batching/pending token \cite{10.5555/3295222.3295349} caps) and GPU capacity-aware routing, which correlated with measurable reductions in HTTP 408 rates and improved SLA. Quantitative breakdowns are summarized in \autoref{tab:analysis-summary} (A3) and headline distributions in \autoref{tab:headline-stats}. The method’s granularity thus directly supported engine improvements and provides a template for ongoing engine level resilience work.

\subsection{Operational failures}
Operational failures, as captured by the taxonomy, accounted for $\approx$4\% of incidents (\autoref{tab:analysis-summary} A4), often coinciding with rollout windows or alerting gaps. By distinguishing these from infrastructure and engine issues, \textbf{the method highlighted the need for adaptive monitoring and deployment validation}. Gaps in observability (low-traffic endpoints, permissive thresholds) and traffic routing (imbalances) delayed detection and prolonged impact. The latter is evident in \autoref{ttm}. \textbf{The taxonomy’s operational lens enabled targeted process refinements} such as dynamic thresholds and improved alerting which are reflected in reduced detection and mitigation latencies. This demonstrates the taxonomy’s value in driving both technical and procedural improvements.

\begin{table}[ht]
\centering
\caption{Incident analysis by root cause category with per category totals and within category breakdowns (Apr-Jun 2025)}
\label{tab:analysis-summary}
\small
\renewcommand{\arraystretch}{1.2}
\setlength{\tabcolsep}{6pt} 
\begin{threeparttable}
\begin{tabular}{p{0.01\columnwidth} p{0.2\columnwidth} p{0.07\columnwidth} p{0.07\columnwidth} p{0.45\columnwidth}}
\toprule
\textbf{ID} & \textbf{Category} & \textbf{Total $n/N$} & \textbf{\% of all} & \textbf{Breakdown (within category)}\tnote{1} \\
\midrule
A1 & \textbf{Infrastructure failures\tnote{2}}
   & 31/156
   & $\approx$20\%
   & Misconfigured deployment templates: $\approx$26\%; Bad nodes: $\approx$19\%; MPI \cite{mpi41} ring: $\approx$18\%; Sidecar: $\approx$11\%; Istio \cite{istio}: $\approx$7\%; Delayed compute: $\approx$7\%; Redis \cite{redis-github}: $\approx$4\%; Storage: $\approx$4\%; Scheduling: $\approx$4\%.\\   
\addlinespace[0.25em]
\midrule
A2 & \textbf{Model configuration failures\tnote{3}}
   & 25/156
   & $\approx$16\%
   & Header misconfig/omission: $\approx$81\%; Invalid outputs due to misconfig: $\approx$19\%.\\
\addlinespace[0.25em]
\midrule
A3 & \textbf{Inference engine failures\tnote{4}}
   & 94/156 
   & $\approx$60\%
   & Timeouts: $\approx$40\%; Resource exhaustion: $\approx$29\%; Crashes: $\approx$12\%; Latency spikes: $\approx$6\%; Input fetch: $\approx$6\%; KV cache: $\approx$5\%; Constraint sampling: $\approx$1\%; Memory leaks: $\approx$1\%.\\
\addlinespace[0.25em]
\midrule
A4 & \textbf{Operational failures\tnote{5}}
   & 6/156 
   & $\approx$4\%
   & During rollout windows: $\approx$33\%; low-traffic: $\approx$17\%; Misconfigured alerts: $\approx$17\%; Error-code translation: $\approx$17\%; Imbalanced traffic routing: $\approx$16\%.\\
\addlinespace[0.25em]
\bottomrule
\end{tabular}
\begin{tablenotes}
\item[1] All percentages sum to 100\% per statistics grouping.
\item[2] Infrastructure failure implications: Auto node restart \cite{kubernetes_cmd}; enable rapid node rebalancing \cite{10.1145/2890784}; buffer GPU strategy where feasible.
\item[3] Model configuration failure implications: Standardize model header schema; add pipeline checks to cut rollout HTTP 5xx errors.
\item[4] Inference engine failure implications: Use connection liveness strategies for long-running requests (streaming, request batching/pending token \cite{10.5555/3295222.3295349} caps), GPU capacity-aware routing.
\item[5] Operational failure implications: Strengthen deployment validation and adaptive alerting; improve traffic routing.
\end{tablenotes}
\end{threeparttable}
\end{table}

 \section{MITIGATION analysis}
 \label{sec:MITIGATION ANALYSIS}

\begin{figure}[ht]
\centering
\includegraphics[width=\columnwidth]{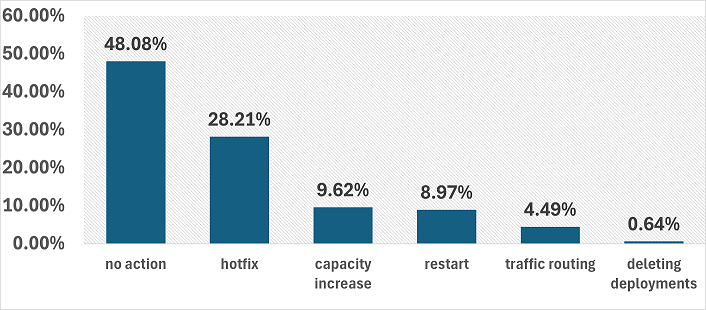}
\caption{Mitigation action mix (percentage)}
\Description{}
\label{mitigation_types}
\end{figure}

 Building on the incident analysis in Section 4, we synthesize the mitigation actions we applied and their operational effect. \autoref{tab:mitigation-summary} summarizes mitigation actions (M1–M6), their shares, usage and automation. \autoref{autodetect} and \autoref{mitigation_types} show the automation split and mitigation action mix. We then discuss our learnings such as AIOps automation gaps (Section 5.1), intelligent failover effectiveness (Section 5.2) before analysing lifecycle metrics (Section 5.3). \\\\
 We define incident lifecycle metrics as \textbf{TTD} (\textbf{Detection}: onset $\rightarrow$detection latency), \textbf{TTE} (\textbf{Diagnosis}: detection$\rightarrow$diagnose latency), and \textbf{TTM} (\textbf{Mitigation}: onset $\rightarrow$mitigation latency), mirroring the definitions in Section 3.4. To complement the incident level metrics, we also track inference request lifecycle metrics \cite{agrawal2024tamingthroughputlatencytradeoffllm} \textbf{TTFT} (request$\rightarrow$first token), \textbf{TBT} (inter-token latency), \textbf{TTLT} (request$\rightarrow$final token), and \textbf{HTTP error rates/status codes}.  Finally in Section 5.4 \autoref{tab:cause-mitigation-outcome} maps Cause $\rightarrow$ Mitigation $\rightarrow$ Outcome for the most common failure patterns and preventative service level controls. 

\begin{table}[ht]
\centering
\caption{Mitigation actions, automation level, and observed operational effects (Apr-Jun 2025)}
\label{tab:mitigation-summary}
\small
\renewcommand{\arraystretch}{1.2}
\setlength{\tabcolsep}{6pt} 
\begin{threeparttable}
\begin{tabular}{p{0.01\columnwidth} p{0.3\columnwidth} p{0.25\columnwidth} p{0.1\columnwidth} p{0.1\columnwidth}}
\toprule
\textbf{ID} & \textbf{Mitigation action} & \textbf{Applied how} & \textbf{Total $n/N$} & \textbf{\% of all}\tnote{6} \\
\midrule
\multicolumn{5}{l}{\textbf{Impactful but expensive}} \\
M1 & \textbf{Hotfix\tnote{1}} & Manual (code/config change) & 44/156 & $\approx$28.21\%  \\
M2 & \textbf{Capacity increase\tnote{2}} & Manual (scale out) & 15/156 & $\approx$9.62\% \\
\midrule
\multicolumn{5}{l}{\textbf{Rapid recovery actions}} \\
M3 & \textbf{Traffic routing (intelligent failover)\tnote{3}} & Partial automation (policy + health probes) & 7/156 & $\approx$4.49\% \\
M4 & \textbf{Restart (node rebalancing)\tnote{4}} & Partial automation (infra playbooks) & 14/156 & $\approx$8.97\%  \\
\midrule
\multicolumn{5}{l}{\textbf{Automation gap}} \\
M5 & \textbf{No action/monitor only\tnote{5}} & Manual decision & 75/156 & $\approx$48.08\% \\
M6 & \textbf{Deleting deployments} & Manual decision & 1/156 & $\approx$0.64\% \\
\addlinespace[0.35em]
\bottomrule
\end{tabular}
\begin{tablenotes}
\item[1] Highest TTM due to validation cost; used when automation cannot resolve root cause. Related to \autoref{ttm}. 
\item[2] Reduced overload/timeout symptoms.  
\item[3] Shortens impact window; reroutes to healthy endpoints/regions.
\item[4] Faster recovery for bad nodes/MPI ring events.
\item[5] Used when transient or self-healing; still consumed manual cycles (part of “resolved without hotfix”).
\item[6] All percentages sum to 100\%; Mitigation actions are coded as mutually exclusive per incident using the first effective action rule. Related to \autoref{autodetect} and \autoref{mitigation_types}.
\end{tablenotes}
\end{threeparttable}
\end{table}

\subsection{AIOps}
\label{subsec:AIOps}
\begin{figure}[ht]
\centering
\includegraphics[width=\columnwidth]{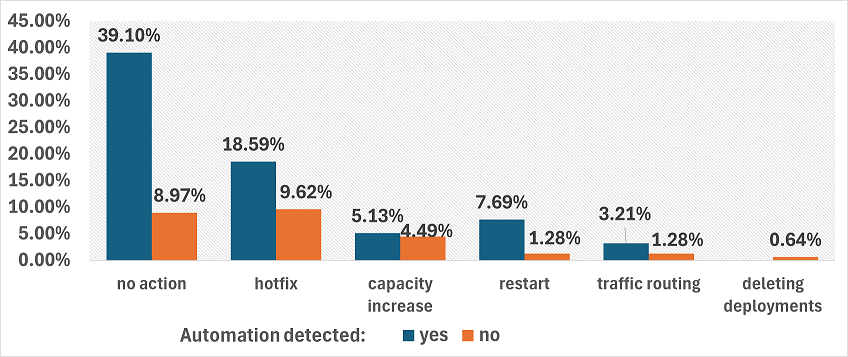}
\caption{Automatic detection status by mitigation action}
\Description{}
\label{autodetect}
\end{figure}

LLM assisted DRI tooling (ICM Analyzer, DRI Copilot) and enrichment pipelines improved triage and reduced manual toil, but did not eliminate the need for human decisions in most cases. We define auto-detected as alerts fired by health probes/DRI pipelines before customer reporting. As observed in \autoref{autodetect}, across the 156 incidents, 115 were auto-detected ($\approx$74\%), and the rest were still not auto-detected. \autoref{tab:mitigation-summary} indicates that high impact mitigations (e.g., hotfix, capacity increase) remained largely manual. \textbf{The implication is to further automate} traffic routing, restart, and capacity increase decisions and to expand pre‑deployment validation (model header schema checks) so that fewer incidents require code fixes (\autoref{tab:mitigation-summary} M1) or ad hoc capacity increase (\autoref{tab:mitigation-summary} M2). 

\subsection{Intelligent failover}
\label{subsec:Intelligent failover}
Traffic routing and node rebalancing (\autoref{tab:mitigation-summary} M3, M4) \textbf{reduced impact windows by shifting traffic off degraded endpoints, while using connection liveness strategies  reduced HTTP 408s in long-running requests}. We treated connection liveness rollouts as \autoref{tab:mitigation-summary} M1 when delivered as code/config changes. These actions are especially effective when combined with accurate health probes and GPU capacity-aware routing policies. As shown in \autoref{autodetect} and \autoref{mitigation_types}, non‑hotfix actions dominate overall (including ‘monitor only’ cases), though most still require manual judgment; expanding automation should reduce TTM.\\

\begin{figure}[ht]
\centering
\includegraphics[width=\columnwidth]{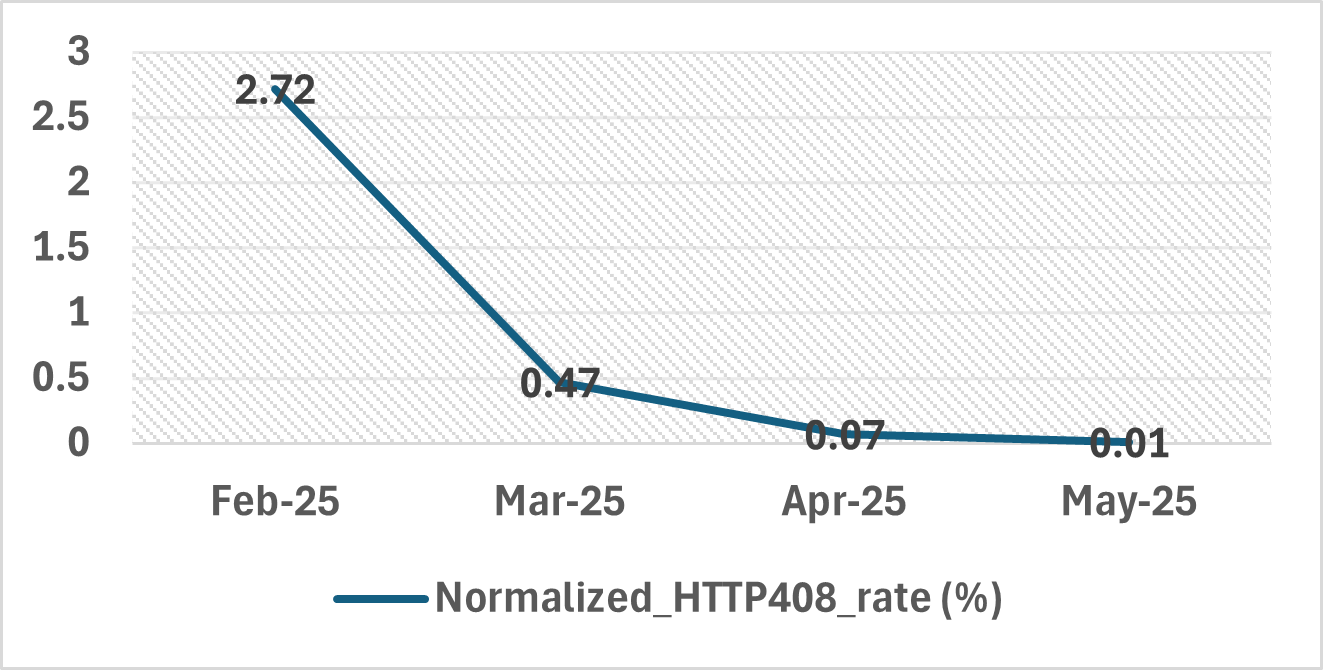}
\caption{Normalized HTTP 408 rate by month (February–March 2025; shown for illustrative purposes, outside the main study window)}
\Description{}
\label{timeouts_month}
\end{figure}

\begin{figure}[ht]
\centering
\includegraphics[width=\columnwidth]{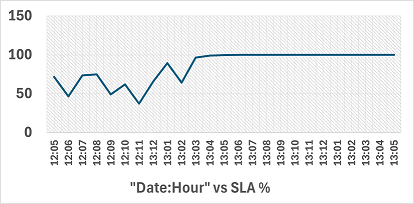}
\caption{SLA improvement after failover to different endpoint}
\Description{}
\label{sla_fix}
\end{figure}

\noindent For example, from \autoref{timeouts_month}, we observe that for the selected impacted customer, the normalized HTTP 408 rate (calculated as the percentage of 408 errors divided by total requests) declined from 2.72\% to 0.47\% (February $\rightarrow$ March 2025), suggesting a directional association between the connection liveness mitigation and the reduction in timeout errors. Note: While our main quantitative analysis focuses on incidents from Apr-Jun 2025, \autoref{timeouts_month} presents an illustrative pre-window trend to highlight the effect of mitigation strategies. Also to clearly illustrate the trend, we present normalized data from one customer, although similar improvements were observed across others where the mitigation was rolled out. Further, SLA is calculated as the percentage of successful requests over total valid requests, excluding client-side issues. In another example (\autoref{sla_fix}), we observed that failover to a different region or endpoint resulted in better SLA with lower TTM. \textbf{Due to industry confidentiality agreements, raw incident data cannot be shared; we present only anonymized and aggregated data.} Also see Section 8 for details on attribution bias mitigation.

\subsection{Comparative analysis of lifecycle metrics}
\label{subsec:Comparative analysis of lifecycle metrics}
To evaluate the effectiveness of incident response and identify systemic inefficiencies, we analyze P50 and P75 for the incident lifecycle—\textbf{Detection (TTD)}, \textbf{Diagnosis (TTE)}, and \textbf{Mitigation (TTM)} and for the request lifecycle—\textbf{TTFT}, \textbf{TBT}, \textbf{TTLT}  capturing typical and moderately elevated behavior under normal and mildly stressed conditions; we report \textbf{Postmortem} insights qualitatively. Focusing on the 50th and 75th percentiles surfaces central tendency and early degradation signals that are more actionable in day-to-day operations than extreme outliers. We omit \textbf{TBT} due to confidentiality as its calculation will reveal the generated token count.\\

\begin{figure}[ht]
\centering
\includegraphics[width=2.5in]{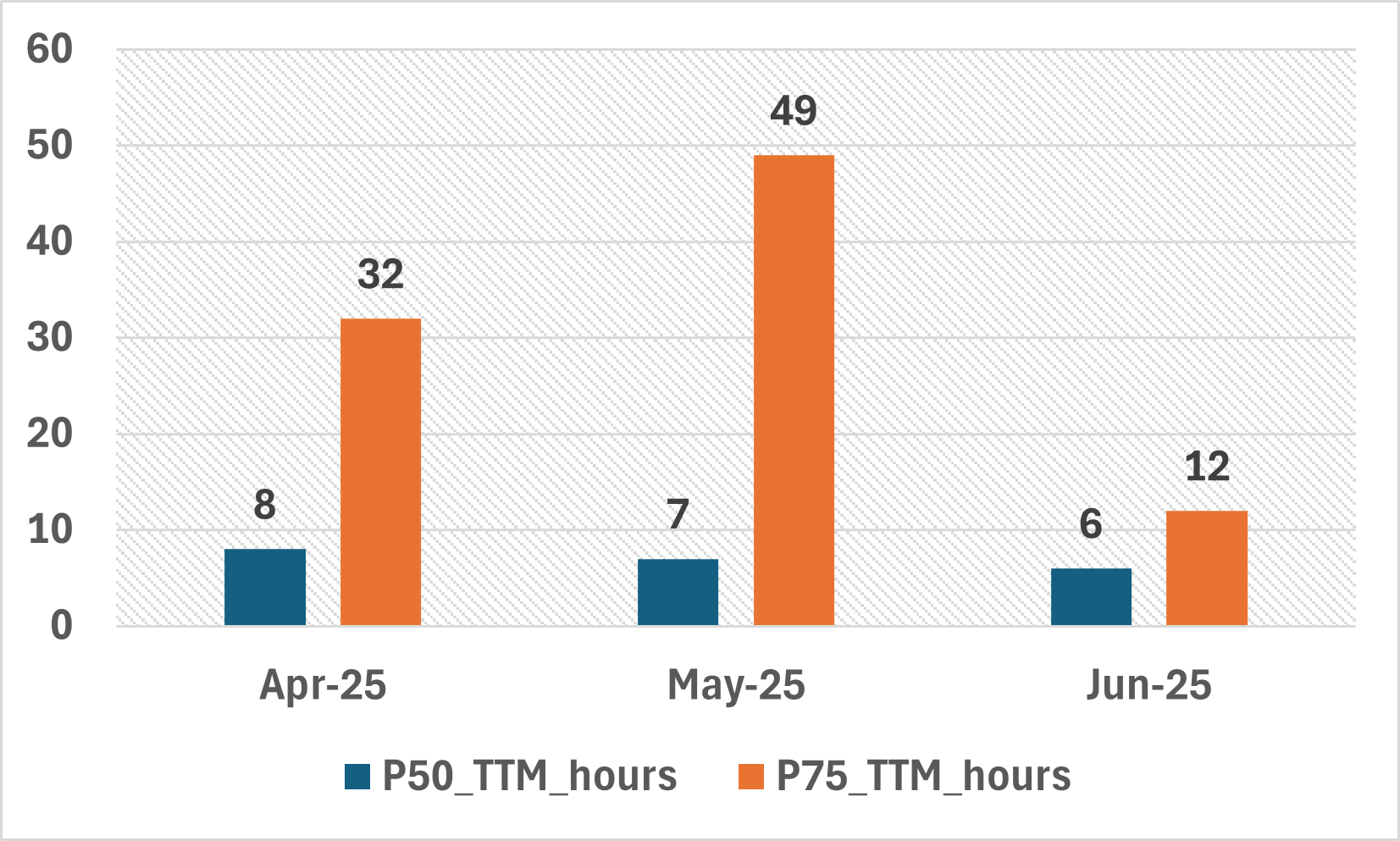}
\caption{Time to Mitigate by month}
\Description{}
\label{ttm}
\end{figure}

\noindent In April 2025, \textbf{P75 TTM} was $\approx32$ hours; in May it spiked to $\approx49$ hours (\autoref{ttm}), reflecting the cost of manual interventions. Over the same window, token generation increased by $\approx1.27\times$ (April$\rightarrow$June). Serving‑side percentiles rose accordingly: \textbf{P75 TTFT} by $\approx60$ ms and \textbf{P75 TTLT} by $\approx80$ ms (\autoref{tab:Inference request lifecycle metrics}), consistent with emerging engine saturation, KV cache inefficiencies, or architectural bottlenecks under increased load.\\

\noindent Mitigation choices manifest in both lifecycle metrics. As \autoref{ttm} indicates,
hotfix-driven periods inflate TTM. In contrast, throughput oriented mitigations (\autoref{tab:mitigation-summary}, M3, M4) e.g., intelligent failover along with GPU capacity-aware routing and connection liveness strategies for long-running requests (\autoref{tab:cause-mitigation-outcome}) are intended to bend the serving‑side curves (\autoref{tab:Inference request lifecycle metrics}). \textbf{We therefore interpret M1, M2 in \autoref{tab:mitigation-summary} as impactful but expensive, and M3, M4 as rapid recovery actions that merit further automation}.

\begin{table}
\centering
\caption{Inference request lifecycle metrics}
\label{tab:Inference request lifecycle metrics}
\footnotesize
\renewcommand{\arraystretch}{1.2}
\setlength{\tabcolsep}{6pt} 
\begin{tabular}{p{0.4\columnwidth} p{0.1\columnwidth} p{0.1\columnwidth} p{0.1\columnwidth}} 
\toprule
         \textbf{Metric in seconds}&  \textbf{Apr 2025}&  \textbf{May 2025}& \textbf{Jun 2025}\\
         \midrule
         P50 TTFT&  0.09&  0.1& 0.1\\
         \midrule
         P75 TTFT&  0.21&  0.27& 0.27\\
         \midrule
         P50 TTLT&  0.25&  0.27& 0.28\\
         \midrule
         P75 TTLT&  0.83&  0.93& 0.91\\
         \bottomrule
\end{tabular}
\end{table}

\subsection{Cause, mitigation, outcome}
\label{subsec:cause_mitigation_outcome}
\autoref{tab:cause-mitigation-outcome} highlights representative incident patterns, mapping each root cause to the corresponding mitigation strategy and observed outcome. This synthesis distills practical lessons from the dataset, illustrating how targeted interventions such as keeping connection liveness for engine timeouts directly improved reliability and reduced operational impact. Internal dashboards confirm the observed directional declines (\autoref{timeouts_month}).

\begin{table}[ht]
    \centering
    \caption{Cause $\rightarrow$ Mitigation $\rightarrow$ Outcome (Selected patterns)}
    \label{tab:cause-mitigation-outcome}
    \centering
    \small
    \renewcommand{\arraystretch}{1.2}
    \setlength{\tabcolsep}{6pt}    
    \begin{threeparttable}
    \begin{tabular}{p{0.3\columnwidth} p{0.35\columnwidth} p{0.25\columnwidth}} 
    \toprule
        \textbf{Cause} & \textbf{Mitigation} & \textbf{Observed outcome} \\
        \midrule
        Bad nodes / MPI ring failures (\autoref{tab:analysis-summary} A1) & Restart nodes; node rebalancing (\autoref{tab:mitigation-summary} M4) & Faster recovery; fewer cascading failures \\
        \midrule
        Model header misconfigurations (\autoref{tab:analysis-summary} A2) & Standardize/validate headers; inner ring rollout checks (\autoref{tab:mitigation-summary} M1, M6) & Fewer HTTP 5xx during deployments\\
        \midrule
        Engine connection timeouts (liveness) under long-running requests (\autoref{tab:analysis-summary} A3) & Enable streaming; tune request batching and cap pending tokens (\autoref{tab:mitigation-summary} M1) & Fewer HTTP 408s; higher success rate \\
        \midrule
        Imbalanced traffic routing (\autoref{tab:analysis-summary} A4) & Intelligent traffic routing; per‑endpoint isolation for high‑priority customers (\autoref{tab:mitigation-summary} M3) & Lower HTTP 429s; reduced scope of impact \\
        \bottomrule
    \end{tabular}
    \end{threeparttable}
\end{table}

\begin{figure}[ht]
\centering
\includegraphics[width=\columnwidth]{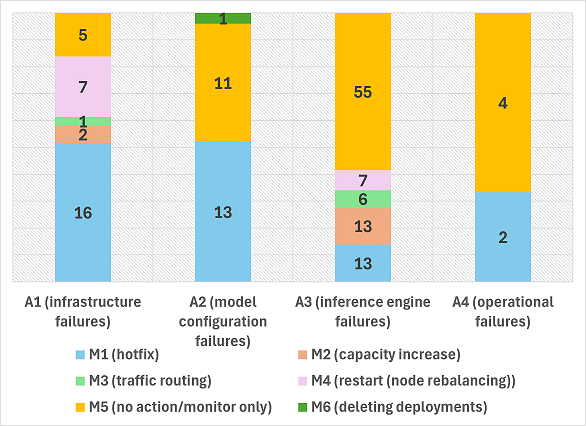}
\caption{Count of mitigation actions (M1-M6) by root cause categories (A1-A4); Column totals equal \autoref{tab:analysis-summary} (A1=31, A2=25, A3=94, A4=6); row totals equal \autoref{tab:mitigation-summary} (M1=44, M2=15, M3=7, M4=14, M5=75, M6=1)}
\Description{}
\label{mitigation_by_category}
\end{figure}

\noindent \autoref{mitigation_by_category} provides an empirical cross-tabulation of root cause categories (A1–A4) by mitigation actions (M1–M6), using the same labeling as \autoref{tab:analysis-summary} and \autoref{tab:mitigation-summary}. This visualization substantiates our Cause$\rightarrow$Mitigation mapping: for example, capacity increase (M2), traffic routing (M3) and restart (M4) are predominantly applied to mitigate inference engine failures (A3), while deleting deployments (M6) is exclusive to model configuration failures (A2). The figure also clarifies the distribution of hotfixes (M1) and monitor-only actions (M5) across all root cause categories, supporting the insights and automation gaps discussed above.

\subsection{Key findings}
\label{subsec:key_findings}
\begin{itemize}
    \item \textbf{Automation gap:} Even with AIOps, a meaningful share of incidents still require manual detection or judgment. Prioritize health probe driven traffic routing, restart, and capacity increase automation. See \autoref{tab:mitigation-summary}, \autoref{autodetect}.
    \item \textbf{Cost of hotfix vs. rapid recovery actions:} Hotfixes restore correctness but extend TTM; traffic routing, and node rebalancing mitigate faster under overload/timeout patterns. See \autoref{tab:mitigation-summary}, \autoref{ttm}.    
\end{itemize}

\section{SELECT CASE STUDIES}
 \label{sec:ADDITIONAL CASE STUDIES}
This section presents a selection of noteworthy case studies that extend and complement the primary findings discussed earlier. Each case offers unique insights or challenges, contributing to a deeper understanding of the subject matter. These examples highlight variations in context, methodology, or outcomes that may inform future research or practice.

\paragraph{Case A: Throughput/capacity pressure in vision \& reasoning models}
\begin{itemize}
    \item \textbf{Cause:} Under bursty load, image encoders and model containers saturated (GPU/CPU imbalance, KV cache pressure) and traffic with unusually high sampling parameters overwhelmed serving, leading to HTTP 429/5xx and timeouts.
    \item \textbf{Mitigation:} We enabled dynamic pushback at the load balancer (caps on pending tokens/requests), tuned request batching and GPU/CPU replica counts, improved gRPC backoff/error mapping, and imposed limits on sampling parameters with per client throttling.
    \item \textbf{Outcome:} Availability stabilized and HTTP 429s were reduced; timeout driven HTTP 5xx decreased under stress. These patterns align with \autoref{tab:cause-mitigation-outcome} (``Imbalanced traffic routing`` and ``Engine connection timeouts/liveness``) and the headline trends in \autoref{tab:headline-stats}.
\end{itemize}

\paragraph{Case B: Timeouts \& response quality (Connection liveness \& output robustness)}
\begin{itemize}
    \item \textbf{Cause:} long-running requests triggered frontend HTTP 408 timeouts despite backends eventually returning HTTP 200s; under load, speech outputs occasionally contained malformed sequences, causing downstream parsing errors (HTTP 5xx).
    \item \textbf{Mitigation:} We used connection liveness strategies like streaming for long-running requests, introduced endpoint level isolation for high‑priority customers, and hardened structured retries across the stack.
    \item \textbf{Outcome:} Intermittent HTTP 408s on long tasks were effectively reduced, and HTTP 5xx due to output anomalies declined. See \autoref{tab:cause-mitigation-outcome} (``Enable streaming``, ``per‑endpoint isolation``).
\end{itemize}

\section{IMPLICATIONS AND FUTURE DIRECTIONS}
 \label{sec:IMPLICATIONS AND FUTURE DIRECTIONS}
 \subsection{Practitioner checklist}
 \label{subsec:prac_checklist}
The below is a practitioner checklist that can be applied to any LLM serving stack. These prescriptions generalize when services exhibit: (i) long-running requests that risk idle timeouts, (ii) diverse model/engine deployments and metadata configurations and (iii) region or endpoint diversity requiring robust health probe driven failover.
\begin{itemize}
    \item \textbf{Adopt a simple incident taxonomy} and periodic labeling reviews to improve triage consistency.
    \item \textbf{Implement health probes, auto restart, and auto traffic routing} for resilient intelligent failover and node rebalancing.    
    \item \textbf{Standardize model headers} with schema validation and pipeline checks to reduce deployment errors.
    \item \textbf{Enable connection liveness for long-running requests} to lower timeout rates and improve response reliability.    
    \item \textbf{Use adaptive monitoring and dynamic thresholds} for low-traffic endpoints to catch silent degradations.    
    \item When sharing evidence is restricted, \textbf{report directional trends} and describe the monitoring signals that informed mitigation decisions.
\end{itemize}

\subsection{Future directions}
Building on the evidence in the Incident analysis (\autoref{tab:analysis-summary}), the operational levers in the Mitigation analysis (\autoref{tab:mitigation-summary}), and the prescriptive patterns in Cause$\rightarrow$Mitigation$\rightarrow$Outcome (\autoref{tab:cause-mitigation-outcome}), we outline the following priorities to harden large scale LLM inference services:

\begin{itemize}
  \item \textbf{Infrastructure efficiency and resilience:}
  Prioritize efficient GPU reuse and rapid, reliable node rebalancing to reduce impact from bad nodes and node failures (\autoref{tab:analysis-summary} A1). Operationalize this via probe driven node restart (\autoref{tab:mitigation-summary} M4), as captured in (\autoref{tab:cause-mitigation-outcome}).

  \item \textbf{Intelligent failover by default:}
  Expand health probe driven failover and GPU capacity-aware routing so traffic autonomously shifts off degraded endpoints (\autoref{tab:mitigation-summary} M3). This directly addresses engine/operational degradation modes (\autoref{tab:analysis-summary} A3, A4) and the imbalanced traffic pattern (\autoref{tab:cause-mitigation-outcome}).

  \item \textbf{Proactive capacity planning and auto remediation:}
  Treat capacity as a first class SLO (Service Level Objective): forecast token/throughput growth and automate capacity increase decisions to avoid manual hotfix/capacity interventions that inflate TTM (\autoref{tab:mitigation-summary} M1, M2). Tie capacity related policies to saturation signals surfaced in request lifecycle trends (\autoref{tab:Inference request lifecycle metrics}).

  \item \textbf{Standardized model header management in the pipeline:}
  Eliminate rollout time configuration faults by enforcing a standard header schema with validation (\autoref{tab:analysis-summary} A2).

  \item \textbf{Engine level resource efficiency under long-running requests and heavy sampling:}
  Optimize serving for long-running requests and high sampling regimes to reduce timeouts/resource exhaustion (\autoref{tab:analysis-summary} A3). Connection liveness strategies and GPU capacity-aware routing are the primary levers. See \autoref{tab:cause-mitigation-outcome}.

  \item \textbf{Adaptive monitoring and richer engine telemetry:}
  Close detection gaps on low-traffic endpoints and ambiguous error translation by deploying adaptive alerting and expanding engine level telemetry (\autoref{tab:analysis-summary} A4). This reduces TTM and improves the effectiveness of intelligent failover (\autoref{tab:mitigation-summary} M3).
\end{itemize}

\noindent Collectively, these directions strengthen resilience, lower operational overhead, and sustain scalability as model complexity and demand rise while aligning day‑to‑day operations with the failure modes and mitigation levers already evidenced in \autoref{tab:analysis-summary}, \autoref{tab:mitigation-summary}, and \autoref{tab:cause-mitigation-outcome}.

\section{THREATS TO VALIDITY}
 \label{sec:THREATS TO VALIDITY}
Manual incident classification may introduce subjectivity, but we mitigated this through iterative cross-validation with engineering experts. Despite efforts to reduce bias, some validity threats remain, which we outline below.

\begin{itemize}
\item \textbf{Scope bias}: While the quantitative analysis is scoped to three months of severity 2 \& 2.5 incidents and lower severity signals may be under represented, the operational practices, taxonomy, and mitigation strategies reflect a year of continuous improvement and learning.
\item \textbf{Evolving system}: Architecture and processes changed during the window; observed patterns may shift over time.
\item \textbf{Platform specificity}: Though our findings reflect the described LLM serving stack; we argue the taxonomy is broadly applicable.
\item \textbf{Documentation fidelity}: Postmortems and telemetry may contain gaps; we cross checked sources and timelines to mitigate mislabeling.
\item \textbf{Overlapping causes}: Some incidents fit multiple categories; we prioritized the earliest/specific root cause per rulebook.
\item \textbf{Attribution bias:} To mitigate attribution bias when measuring a mitigation's outcome, we excluded maintenance windows, major deployment freezes and verified that no concurrent mitigation or capacity expansions overlapped.
\end{itemize}

\noindent While our analysis is grounded in the described incident management and telemetry systems, the taxonomy, labeling methodology, and lifecycle metrics (TTD, TTE, TTM; TTFT, TBT, TTLT) are designed to be platform agnostic. \textbf{Any provider with access to incident records and runtime logs can apply these definitions} using standard log analysis tools. We encourage other organizations to adapt our taxonomy and metric definitions to their own environments for comparative studies.

\section{CONCLUSION}
 \label{sec:CONCLUSION}
This paper presents a practice based reliability study of hyperscale LLM serving system, offering a structured retrospective analysis of high severity incidents in a production environment. The incident taxonomy and methodology introduced here provides a foundation for ongoing reliability engineering efforts and may serve as a reference for other teams managing mission critical AI infrastructure. By applying the taxonomy and quantifying dominant failure modes and surfacing effective mitigations and automation opportunities, we reveal that most incidents in our dataset were resolved through operational actions or self‑healing rather than code changes underscoring the importance of observability, automation, and well defined incident response workflows. Our results emphasize operational excellence in areas such as monitoring, traffic routing, and standardized configuration, alongside inference engine and infrastructure improvements that reduce mitigation time and impact. By correlating detection, diagnosis, and mitigation stages, we identify key findings that may guide practitioners and future research to enhance LLM service resilience. This study demonstrates how systematic, empirically grounded analysis of inference operations can drive more reliable and cost-efficient LLM serving at scale.

\section*{ACKNOWLEDGMENTS}
\label{sec:ACKNOWLEDGMENTS}
We thank Mahsa Mirza and Aishwariya Raman for their contributions to incident dataset collection and labeling. We are grateful to engineering managers Klein Hu, Chris Basoglu and Rakesh Kelkar for their support and for reviewing this paper. We also appreciate the collaboration and transparency of the entire LLM serving system team in sharing incident data, postmortems, and operational insights. Special thanks to the engineers, incident managers (Mukund Sharma and Tushar Santoki), and product leads whose detailed documentation and retrospective analysis made this study possible. Finally, we acknowledge the reviewers and stakeholders whose feedback helped refine the taxonomy and methodology presented in this work. 

\section*{DATA GOVERNANCE AND ETHICS}
\label{sec:DATA GOVERNANCE AND ETHICS}
This study was conducted using only aggregated and operational telemetry data; no customer Personal Identifiable Information (PII) was accessed or analyzed. No prompts, completions, or customer identifiers were processed in this study. All incident records were handled in accordance with internal privacy and security policies where applicable. Generative AI tools (e.g., Copilot) were used for language refinement only in the preparation of this manuscript. All substantive content, analysis, and conclusions are the authors’ own.

\section*{REPRODUCIBILITY \& ARTIFACT AVAILABILITY}
We plan to make available the incident taxonomy and anonymized examples of the labeling rulebook used in this study. Due to organizational confidentiality requirements, no raw logs, operational data, or source code can be provided. The shared materials are intended to facilitate understanding of our methodology and support transparency, while respecting all data governance constraints.

\bibliographystyle{ACM-Reference-Format}
\bibliography{sigconf_preprint}
\end{document}